\documentstyle[12pt,epsf]{article}
\newcommand{\la}[1]{\label{#1}}

\newlength{\numlen}

\newcommand{\be}{\begin{equation}}
\newcommand{\ee}{\end{equation}}
\newcommand{\ba}{\begin{eqnarray}}
\newcommand{\ea}{\end{eqnarray}}
\newcommand{\rmi}[1]{{\mbox{\scriptsize #1}}}
\newcommand{\bi}{\begin{itemize}}
\newcommand{\ei}{\end{itemize}}

\newcommand{\bfb}{\mbox{\bf b}}

\newcommand{\roots}{$\sqrt{s}$}

\newcommand{\that}{\hat t}
\newcommand{\pth}{\partial_{\hat t}}
\newcommand{\peta}{\partial_\eta}
\newcommand{\vbar}{\bar v}

\newcommand{\fr}[2]{{\frac{#1}{#2}}}

\def\lsim{\raise0.3ex\hbox{$<$\kern-0.75em\raise-1.1ex\hbox{$\sim$}}}
\def\gsim{\raise0.3ex\hbox{$>$\kern-0.75em\raise-1.1ex\hbox{$\sim$}}}

\begin{document}

\begin{titlepage}
\begin{flushright}
CERN-TH/97-96\\
JYFL 6/97\\
5 May 1997\\
nucl-th/9705015\\
\end{flushright}
\begin{centering}
\vfill

{\bf
         HYDRODYNAMICS OF NUCLEAR COLLISIONS WITH INITIAL CONDITIONS
FROM PERTURBATIVE QCD
}

 \vspace{0.5cm}
 K.J.Eskola$^{\rm a,b,}$\footnote{kari.eskola@cern.ch},
 K. Kajantie$^{\rm a,b,}$\footnote{keijo.kajantie@cern.ch} and
 P.V. Ruuskanen$^{\rm c,}$\footnote{ruuskanen@jyfl.jyu.fi}

\vspace{1cm}
{\em $^{\rm a}$ CERN/TH, CH-1211 Geneve 23, Switzerland\\}
\vspace{0.3cm}
{\em $^{\rm b}$ Department of Physics,
P.O.Box 9, 00014 University of Helsinki, Finland\\}
 \vspace{0.3cm}
{\em $^{\rm c}$ Department of Physics, University of Jyv\"askyl\"a,
P.O.Box 35, 40351 Jyv\"askyl\"a, Finland\\}

\vspace{1cm}
{\bf Abstract}

\end{centering}

\vspace{0.3cm}\noindent
We compute the longitudinal hydrodynamic flow in ultrarelativistic
heavy ion collisions at $\sqrt{s}$ = 5500 GeV by using boost non-invariant
initial conditions following from perturbative QCD.  The transfer of
entropy and energy from the central region to larger rapidities
caused by boost non-invariance is determined and the associated decrease
in the lifetime of the system is estimated.
\vfill

\end{titlepage}
\section{Introduction}
One simple scenario for treating the behaviour of
QCD matter formed in the central region
(nearly at rest in the center of mass frame)
in ultrarelativistic heavy ion collisions is to neglect transverse
motion and baryon number, to assume that the
initial conditions for longitudinal motion
are longitudinally boost invariant and to assume that the matter expands
isentropically as an ideal fluid. One then obtains the Bjorken
similarity flow \cite{bj}. The purpose of this study is to
take the initial conditions essentially as given by perturbative
QCD \cite{bm}-\cite{ek}. These have two characteristics. Firstly,
they are boost non-invariant, there is no rapidity plateau but a
wide gaussian-like rapidity distribution. Secondly, due to
the remarkable small-$x$ increase of the nucleon structure functions
observed at HERA \cite{heradata} the initial energy densities are quite
large, at LHC energies of \roots = 5.5 TeV almost 1000 GeV/fm$^3$.
This leads to rather long hydrodynamical evolution times ($\approx$
100 times initial thermalisation time) and allows
a boost non-invariant flow to develop.
In the following we shall, in particular,
study the additional longitudinal flow caused by the maxima
of the initial entropy and energy densities at $y=0$ and the
associated transfer of entropy and energy from the $y=0$ region to
larger rapidities.

We limit this study to LHC energies for the following two reasons.
Firstly, we want to study longitudinal 1+1d hydrodynamical effects and
at LHC, due to the very large initial temperatures, 
this period lasts by far the longest. Also, during all this period
the matter remains in the high $T$ plasma phase. 
All the complications associated with the phase
transition and the hadronic phase arise together with the need
to go over to 1+3d expansion \cite{ruoskanenpuola}-\cite{rischke}.
Within the 1+1d approximation these late-time features have
been studied in, say, \cite{vjmr}.
Secondly, at LHC energies the perturbative computation of the initial
conditions is more reliable.
At RHIC energies of \roots = 200 GeV there is still a
sizable soft component present, which makes the buildup of the initial
energy density slower and would require separate modeling \cite{eg}.
This is even more so at the SPS energies of \roots = 20 GeV, where
boost non-invariant initial conditions are modelled by using
rapidity distributions of final state particles \cite{sollfrank}
or by forcing the flow to be boost invariant \cite{sx}.

We also emphasize that the main aim is to study the evolution of the
flow under a set of approximations which will need corrections when
applied to the physical situation. These include full thermalisation
and validity of given initial conditions at very large rapidities.
Also fluctuations in the initial conditions \cite{grz} and their
variation with transverse coordinate will have to be taken into
account.

\section{The equations}
For the ultrarelativistic 1+1--dimensional similarity flow it is convenient
to replace $x^\mu=(t,x)$ by the proper time $\tau$ and the space-time
rapidity $\eta$:
\ba
\tau&=&\sqrt{t^2-x^2},\quad \eta=\fr12\log{t+x\over t-x},\\
t&=&\tau\cosh\eta,\quad x=\tau\sinh\eta.
\ea
The variable
\be
\hat t=\log(\tau/\tau_i)
\ee
also naturally appears. The general equation of state with one
conserved quantum number is
\ba
p&=&p(T,\mu), \\
s={\partial p\over\partial T},\quad n_q-n_{\bar q}&\equiv& n=3n_B=
{\partial p\over\partial\mu},\quad
\epsilon=Ts-p+\mu n.
\label{thermo}
\ea
The aim now is to determine, for given initial conditions,
the pressure $p(t,x)$ and the flow
$v(t,x)\equiv\tanh\Theta(t,x)$ from the hydrodynamic equations
\be
\partial_\mu T^{\mu\nu}=0,\quad\nu=0,1,\qquad\partial_\mu J_B^\mu=0,
\label{eom}
\ee
where
\ba
T^{\mu\nu}&=&(\epsilon+p)u^\mu u^\nu-pg^{\mu\nu}, \quad
J_B^\mu=n_Bu^\mu,\label{emtens}\\
u^\mu&=&(\gamma,\gamma v)=(\cosh\Theta,\sinh\Theta).
\ea
From eqs. (\ref{emtens}) and (\ref{thermo}) it follows that
$u_\nu\partial_\mu T^{\mu\nu}-3\mu\partial_\mu J_B^\mu=
T\partial_\mu s^\mu$, where $s^\mu=su^\mu$ is the entropy
current, so that these equations imply entropy conservation:
\be
\partial_\mu s^\mu=0,\quad s^\mu=su^\mu. \label{scons}
\ee

To express eqs.\ (\ref{eom}) in component form it is convenient
to take their components parallel ($u_\nu\partial_\mu T^{\mu\nu}=0$)
and orthogonal
(($g_{\alpha\nu}-u_\alpha u_\nu)\partial_\mu T^{\mu\nu}=0,\,\,
\alpha=0,1$) to $u_\mu$. The equations to be solved then become
\cite{krr}
\ba
(\pth+\vbar\peta)\epsilon+(\epsilon+p)(\vbar\pth+\peta)\Theta&=&0,\\
\label{eq1}
(\vbar\pth+\peta)p+(\epsilon+p)(\pth+\vbar\peta)\Theta&=&0,\\
\label{eq2}
(\pth+\vbar\peta)n_B+n_B(\vbar\pth+\peta)\Theta&=&0,\la{eq3}
\ea
where
\be
\vbar(\tau,\eta)=\tanh[\Theta(\tau,\eta)-\eta)],
\ee
with given initial conditions $T=T(\tau_i,\eta),\,\,\mu=
\mu(\tau_i,\eta)$.

\begin{figure}[tb]
\epsfysize=10cm
\centerline{\epsffile{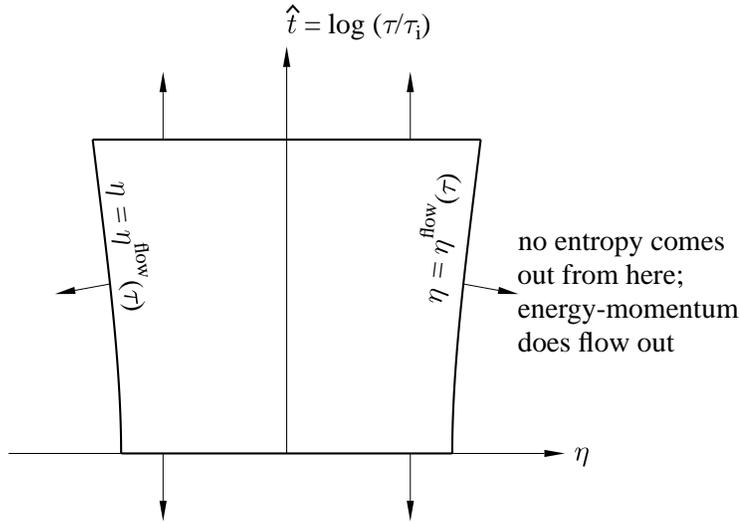}}
\vspace*{-5.5cm}
\caption[a]{\small Section of space-time between two flow lines
(eq.~(\ref{flowdef})) and two lines of constant proper time.}
\la{flowlines}
\end{figure}

To correctly interpret the numerical results it is useful to have
a concrete picture of the role played by the conservation laws in
the flow. This is obtained by writing for any conserved vector $V^\mu$
\be
0=\int d^2x\,\partial_\mu V^\mu=\int_C\,d\sigma_\mu V^\mu,
\la{cons}
\ee
with $dx^\mu=(dt,dx)$ and $d\sigma^\mu=(dx,dt)$ and choosing
the path as shown in
Fig.~\ref{flowlines}. Here the horizontal
lines are two lines of constant $\tau$ while the
vertical lines are chosen as flow lines $\eta=\eta^\rmi{flow}(\tau)$,
defined as the solutions of
\be
{dx(t)\over dt}= v(x(t),t) \Rightarrow \tau{d\eta\over d\tau}=
\vbar(\tau,\eta(\tau))=\tanh[\Theta(\tau,\eta)-\eta)].
\la{flowdef}
\ee
Computing the line integral (\ref{cons}) over various portions of the
path in Fig.~\ref{flowlines}
gives the fluxes of $V^\mu$ through these
portions; their total sum has to vanish.

Take first $V^\mu=s^\mu$. Converting the line integral in
eq.~(\ref{cons}) to the variables $\tau,\eta$ using
$u^\mu=(\cosh\Theta, \sinh\Theta)$ one finds that
\be
\int_C\,d\sigma_\mu s^\mu  = \int_C\tau\cosh(\Theta-\eta)s\,\,
[d\eta-\tanh(\Theta-\eta)d\tau/\tau].
\ee
Thus the entropy flux through a flow line (\ref{flowdef})
is zero ($d\sigma_\mu u^\mu =0$ is an equivalent definition of a flow
line) while the flow through a line $\tau$ = constant is
\be
 S(\tau,\eta_1<\eta<\eta_2)=\int_{\eta_1}^{\eta_2} d\eta\, \tau
s(\tau,\eta)\cosh[\Theta(\tau,\eta)-\eta].
\la{stot}
\ee
Varying now $\tau$ and letting $\eta_1(\tau),\eta_2(\tau)$
follow flow lines, eq.~(\ref{cons})
implies that the integral (\ref{stot}), the total
entropy measured between two flow lines at a fixed proper time, is
constant, independent of $\tau$. A similar equation holds for $J_B^\mu$.

We may also apply eq.~(\ref{cons}) to the energy and momentum
fluxes $T^{0\mu}$ and $T^{1\mu}$. Since they are not proportional to
$u^\mu$, the result becomes more complicated. A similar 
computation gives that the flux through a flow line is:
\ba
F_E(\tau_1<\tau<\tau_2)&=& -\int_{\tau_1}^{\tau_2} d\tau{p\sinh\Theta\over
\cosh[\Theta-\eta^\rmi{flow}(\tau)]},\la{side0}\\
F_P(\tau_1<\tau<\tau_2)&=& -\int_{\tau_1}^{\tau_2} d\tau{p\cosh\Theta\over
\cosh[\Theta-\eta^\rmi{flow}(\tau)]},\la{side1}
\ea
where the arguments of $\Theta,p$ are $\tau,\eta^\rmi{flow}(\tau)$.
Further, the energy and momentum fluxes through a segment
$\tau$ = constant are:
\ba
E(\tau,\eta_1<\eta<\eta_2)&=& \int_{\eta_1}^{\eta_2}d\eta\,\tau[\epsilon\cosh\Theta
\cosh(\Theta-\eta)+p\sinh\Theta\sinh(\Theta-\eta)],\la{bottom0}\\
P(\tau,\eta_1<\eta<\eta_2)&=& \int_{\eta_1}^{\eta_2}d\eta\,\tau[\epsilon\sinh\Theta
\cosh(\Theta-\eta)+p\cosh\Theta\sinh(\Theta-\eta)],\la{bottom1}
\ea
where the arguments of $\epsilon,p,\Theta$ are $\tau,\eta$. Referring to
Fig.~\ref{flowlines}, the total sum of two contributions of type
(\ref{side0}) and of two contributions of type (\ref{bottom0})
has to vanish for $T^{0\mu}$; similarly for $T^{1\mu}$. This implies
that the energy (eq.~(\ref{bottom0})) or momentum (eq.~(\ref{bottom1}))
between two flow lines is not constant but changes due to energy
or momentum flow across the flow line: work done against expansion.

Note the different flow-dependent factors in (\ref{stot}) and
(\ref{bottom0}): the total energy $E$ contains an additional
boost factor $\cosh\Theta$ not present for $S$.

To have a still simpler view of the conservation laws,
assume a similarity flow,
$\Theta(\tau,\eta)=\eta$ or $v(t,x)=x/t$
and an equation of state $p=p(\epsilon)$.
Then $\bar v=0$ and from eq.~(\ref{flowdef}) the flow lines
are $\eta$ = constant
and we take them to be $\pm\eta_0$. For this very special
flow the equations of motion (\ref{eom}) become
\ba
\tau\partial_\tau\epsilon+\epsilon+p&=&0,\la{simeps}\\
\partial_\eta p&=&0, \\
\tau\partial_\tau n+n&=&0,\\
\tau\partial_\tau s+s&=&0,\la{sims}
\ea
(the last follows from the first and third and $\epsilon+p=
Ts+\mu n$) and further imply that
\be
p=p(\tau),\,\,\epsilon=\epsilon(\tau),\,\,s=s(\tau,\eta)=
{\tau_i\over\tau}s(\tau_i,\eta),\,\,n=n(\tau,\eta)=
{\tau_i\over\tau}n(\tau_i,\eta),
\ee
i.e. $p,\epsilon$ depend on $\tau$ only while $s,n$ can
depend also on $\eta$ in such a way that $Ts+\mu n$ depends only on $\tau$.
The conservation laws of $s^\mu,T^{0\mu}$ then simplify to
\ba
\int_{-\eta_0}^{\eta_0}d\eta\,\,[\tau_1s(\tau_1,\eta)-
\tau_2s(\tau_2,\eta)]&=&0, \la{sconst}\\
\int_{-\eta_0}^{\eta_0}d\eta\,\cosh\eta\,\,[
\tau_1\epsilon(\tau_1)-\tau_2\epsilon(\tau_2)]
&=&2\int_{\tau_1}^{\tau_2}d\tau\,p(\tau)\sinh\eta_0;\la{ep}
\ea
similar ones hold for $J_B^\mu, T^{1\mu}$. Eq.(\ref{sconst})
is clearly an integrated form of eq.(\ref{sims}). In
eq.(\ref{ep}) the $\eta$ dependent parts factor and match on
two sides of the equation. The $\tau$ dependent part explicitly
shows how the change in $\epsilon(\tau)\tau$ is related to
$pdV\sim pd\tau$.

A further step of simplification would be to assume a massless
equation of state, $p=\epsilon/3=aT^4+b\mu^2T^2+c\mu^4$,
$a,b,c$ = constants. Then a solution with a similarity flow
would be
\be
T(\tau,\eta)=\left({\tau_i\over\tau}\right)^{1/3}T(\tau_i,\eta),
\quad
\mu(\tau,\eta)=\left({\tau_i\over\tau}\right)^{1/3}\mu(\tau_i,\eta),
\quad
p(\tau)=\left({\tau_i\over\tau}\right)^{4/3}p(\tau_i),
\ee
where $T(\tau_i,\eta),\mu(\tau_i,\eta)$ are so constrained that
$p(\tau_i)$ is independent of $\eta$. This is the standard Bjorken
flow \cite{bj} generalised by the inclusion of baryon number.
Putting $\mu=0$ finally gives the Bjorken flow.

\section{Initial conditions}
One expects the initial net baryon number to very small near
$y=0$; in \cite{ek} the net baryon number-to-entropy ratio, $B\slash S$
was computed to be about
1/5000. Hence, we shall put the chemical potential $\mu=0$. Further, we are
interested in the evolution of the system in the plasma phase,
$T = T_c ... 7T_c$ and
choose the equation of state to be $p=p(T)=\epsilon/3=aT^4$,
$a$ = constant, $s=p'(T)=(\epsilon+p)/T$. According to lattice data
for pure SU(3) \cite{laermann} the validity of $p=\epsilon/3$ improves with increasing
$T$ so that the error is 20\% at $T=2T_c$ and 10\% at $T=3T_c$.

\begin{figure}[tb]
\vspace*{2cm}
\centerline{\hspace*{-2cm}
\epsfxsize=10cm\epsfbox{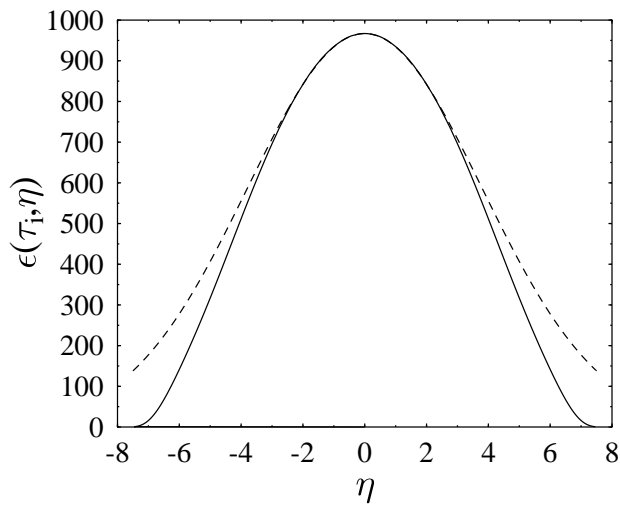}}
\vspace*{-6cm}
\caption[a]{\small The initial condition for $\epsilon(\tau_i,\eta)$ for
 $\sqrt{s}$ = 5500 GeV. The dashed curve is the gaussian fit,
eq.~(\ref{gaussianfit}), with $\sigma=3.8$.}
\la{epsi}
\end{figure}

The initial conditions have to specify $\epsilon(\tau_i,\eta)$
and $\Theta(\tau_i,\eta)$. For the initial flow we shall simply take
a similarity flow, $\Theta(\tau_i,\eta)=\eta$, in order to study
how boost non-invariance affects it.
The initial conditions for the energy density are computed by extending the
calculations of \cite{ekr}-\cite{ek} to all rapidities. The initial energy
per unit rapidity in the local rest frame equals the transverse energy of
produced minijets computed from
\be
\epsilon(\tau_i,\eta)={dE \over d\eta}{1\over V}
\approx T_\rmi{AA}(\bfb=0)\int_{p_0}^\infty dp_T\,p_T
{d\sigma^\rmi{NN}\over dp_Tdy}\cdot{1\over V},
\la{inicond}
\ee
where the nuclear overlap function is $T_\rmi{AA}(0)\approx
A^2/(\pi R_A^2)\approx$ 32/mb for Pb+Pb, the volume per unit rapidity
is $V=\tau_i\pi R_A^2$, $1/\tau_i=p_0=2$ GeV and
the inclusive gluon jet production cross section in NN collisions is
computed in \cite{ek}. The results for $\tau_i=0.1$ fm are
shown in fig.~\ref{epsi} for LHC
energies, \roots = 5500 GeV, together with a gaussian fit to the central
region
 \be
\epsilon(\tau_i,\eta)=\epsilon_0\exp[-\eta^2/(2\sigma^2)],
\la{gaussianfit}
\ee
with $\sigma=3.8$. The distribution is seen to be
quite broad. In this sense boost non-invariance is quite mild
in the central region, even though evidently the initial condition
has to drop faster than a gaussian at the ends of phase space.
As a side remark, at RHIC energy \roots = 200 GeV the width parameter
is $\sigma=2.05$.

In the initial values of $\epsilon_i$ in fig.~\ref{epsi} only
gluons with $p_T\ge2$ GeV are included. The value
2 GeV corresponds to the saturation limit at LHC energies \cite{ek}
and the
resulting $\epsilon_i$ can be expected to be a good estimate for
the total energy density. At RHIC energies the saturation limit
is lower, only about 1 GeV. This is so low a scale that perturbation
theory becomes unreliable. One could also try to keep the minimum $p_T$
of partons included in the computation at the fixed value of 2 GeV;
at RHIC energies one then should also include a sizable (about 50\%)
soft component if one wants to model the entire event. Modeling the
soft component is certainly possible but we wish to avoid this
phenomenological analysis \cite{eg} in the present work.

According to eq.(\ref{bottom0}) the total initial energy carried by the
flow in the interval $-\eta_0<\eta<\eta_0$ is, 
estimating the transverse area to be $\pi R_A^2$,
\be
E_\rmi{tot}(\tau_i)=\pi R_A^2\int_{-\eta_0}^{\eta_0}\tau_i
d\eta \cosh\eta \,\,
\epsilon(\tau_i,\eta) \approx \int_{p_T>p_0,|y|<\eta_0}d^3p\,\,
E{d\sigma^{NN}\over d^3p}\cdot T_{AA}(0).
\la{Etot}
\ee
Here the first part is a general relation for a flow $\Theta=\eta$;
the second part shows where it is computed from.
Due to the boost factor $\cosh\eta$, which in the jet computation
corresponds to the weight factor $E$,
most of the energy is carried by flow at large rapidities.
When $\eta_0$ approaches the beam rapidity $y_\rmi{beam}$, the
error in the jet computation grows for two reasons: firstly,
baryon number will be important at large rapidities and
secondly, physics there is not that of independent 2$\to2$ collisions.
However, this error is important  only at very large values
of $\eta$, where the energy density is already small. It thus will
not affect the conclusions of this hydrodynamical study.

\section{Analytic approximations}
We are studying modifications to the Bjorken flow and it is appropriate
to ask whether they in some limit can be described analytically.
Consider the limit $\sigma\gg1$, a broad gaussian (\ref{gaussianfit}).
Writing ($\that=\log(\tau/\tau_i)$)
\ba
\epsilon(\tau,\eta)&=&\epsilon_0\exp\{
-\fr43\that-{\eta^2\over2\sigma^2}+g(\tau)\},\\
\Theta(\tau,\eta)&=&\eta[1+f(\tau)],
\ea
linearising eqs.(\ref{eq1}-\ref{eq2}) under the assumptions
$f,g,\eta^2/\sigma^2\ll1$ and integrating, one obtains
\ba
f(\tau)&=&{3\over8\sigma^2}[1-\exp(-\fr23\that)],\\
g(\tau)&=&-{1\over2\sigma^2}\{\that-\fr32[1-\exp(-\fr23\that)]\}.
\ea
The linearisation $\partial_{\that} f\gg f^2$ demands that
\be
\that\ll 3\log(\fr43\sigma).
\ee
Furthermore, for the flow lines one obtains
\be
\log[\eta^\rmi{flow}(\tau)/\eta_0]=
{3\over8\sigma^2}\{\that-\fr32[1-\exp(-\fr23\that)]\}
\ee

These equations contain the following natural expectations:
\bi
\item The energy density at $\eta\approx0$ decreases faster than in
the Bjorken flow due to energy moving to larger $\eta$. For small
times
\be
\epsilon(\tau)=\epsilon_0\exp[-\fr43(\that+{\that^2\over8\sigma^2}+...)];
\ee
\item With increasing time the flow is accelerated relative to the
similarity flow. For small times
\be
\Theta(\tau,\eta)=\eta(1+{\that\over4\sigma^2}+...);
\ee
\item The flow lines bend outwards. For small times
\be
\eta^\rmi{flow}(\tau)=\eta_0(1+{\that^2\over8\sigma^2}+...).
\ee

\ei

\begin{figure}[tb]
\vspace*{2cm}
\centerline{\hspace{-1.8cm}
\epsfxsize=10cm\epsfbox{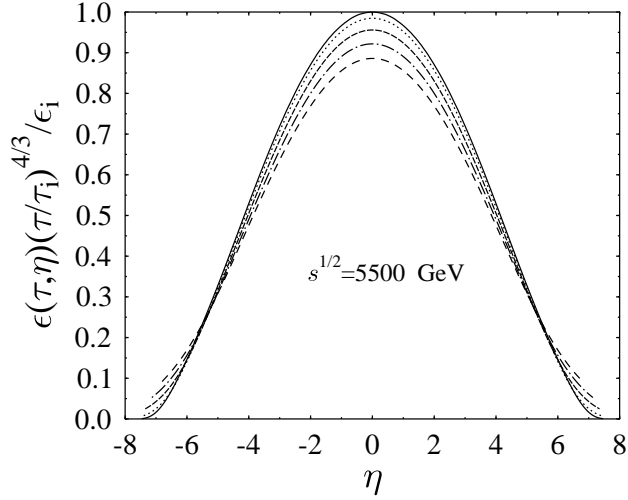}}
\vspace*{-6cm}
\caption[a]{{\small The energy density scaled by the Bjorken flow,
$\epsilon(\tau,\eta)/\epsilon_i*(\tau/\tau_i)^{4/3}$, for
 $\sqrt{s}$ = 5500 GeV. From top the curves correspond to
$\tau/\tau_i=1,10^{1/2},10,10^{3/2},100$.}}
\la{eps}
\end{figure}
\begin{figure}[tb]
\vspace*{2cm}
\centerline{\hspace{-1.8cm}
\epsfxsize=10cm\epsfbox{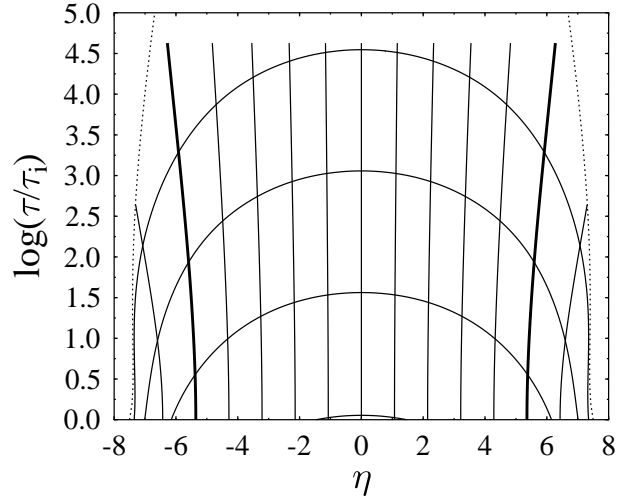}}
\vspace*{-6cm}
\caption[a]{{\small Curves of constant $\epsilon(\tau,\eta)$ for
 $\sqrt{s}$ = 5500 GeV. The curves from below at $\eta=0$
correspond to $\epsilon$ = 900,117,15,2 GeV/fm$^3$. The vertical lines
are flow lines ($\eta$ = constant for Bjorken flow). The thick flow
lines are used to compute fig.~\ref{s}. The dotted lines are the
leftmost characteristic curve $C^+$ and the rightmost $C^-$, between
which the computation has to remain.}}
\la{epsf}
\end{figure}

\section{Numerical results}
The method of characteristics \cite{collatz} is particularly suited
for 1+1 dimensional hydrodynamical problems. First the differential
equation \cite{krr}
\be
{d\eta\over d\that}=\tanh[\Theta(\that,\eta)-\eta\pm y_s],
\la{chareq}
\ee
where $y_s=\tanh c_s=\tanh(1/\sqrt3)$, defines the two families
$C^\pm$ of characteristics. This is like the equation (\ref{flowdef})
defining the flowlines, but modified by the sound rapidity
$\pm y_s$. Along each family of curves the changes of $\Theta$
and $\epsilon$ are related by
\be
d\Theta\pm{c_s\over \epsilon+p}=0,\qquad {\rm along}\,\,C^\pm.
\la{along}
\ee
A code integrating $\epsilon,\Theta$, starting from the initial values
$\epsilon(\tau_i,\eta),\Theta(\tau_i,\eta)=\eta$ by determining the
characteristic directions, and then the new values of $\epsilon,
\Theta$ by stepping in the characteristic directions, is easily
constructed.

Results of numerical integration of the equations are shown in
Figs.~\ref{eps}-\ref{s} for LHC (\roots = 5500 GeV).
The energy density, scaled by the Bjorken flow, is shown in fig.~\ref{eps}
as a function of $\eta$ at different
times. Curves of constant energy density on
the $\tau,\eta$ plane are plotted in fig.~\ref{epsf}.
The energy density in fig.~\ref{eps} at $\eta=0$ is seen to decrease
somewhat faster than in the Bjorken case.
This is due to leakage of energy to
larger rapidities, seen as an increase in $\epsilon$ at large $\eta$
and described analytically in section 4.
During the whole duration of the plasma phase,
$\epsilon\gsim2$ GeV/fm$^3$, or from $\tau_i$ to about 100$\tau_i$,
this additional decrease is about 12\% at LHC.

Converted to the lifetime of the system in the plasma phase
(the time it takes to decrease from $\epsilon_i$ to $\epsilon_c$)
the above result implies that the density gradient in the
longitudinal direction decreases the lifetime by about 9\%.
This is opposite to the effects of dissipation:
in the case of a rapidity plateau, the fastest decrease of energy
density is obtained in the case of full thermalisation; dissipative
effects increase the lifetime of the system \cite{eg}. 
The longest lifetime is obtained for free-streaming expansion. 

Fig.~\ref{epsf} also shows flow lines, which are constant in $\eta$
in the
Bjorken case but bend outwards in the present case. We do not show the
numerical grid of characteristic curves, but the
leftmost characteristic curve $C^+$ and the rightmost one $C^-$ are plotted
in this figure. The numerical computation
cannot be extended outside them.

\begin{figure}[tb]
\vspace*{2cm}
\centerline{\hspace{-1.8cm}
\epsfxsize=10cm\epsfbox{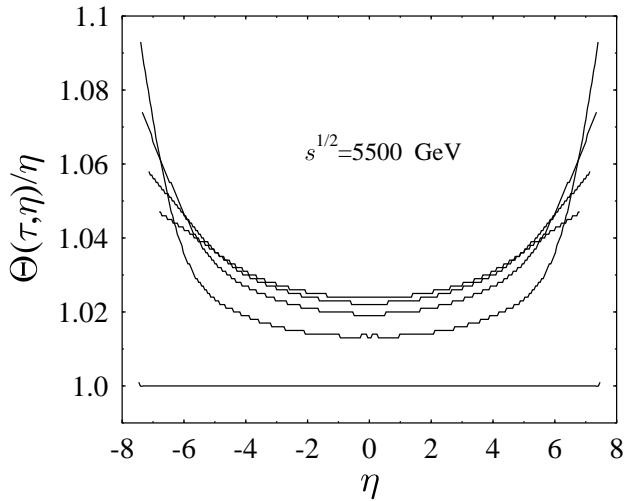}}
\vspace*{-6cm}
\caption[a]{{\small The flow rapidity scaled by $\eta$,
$\Theta(\tau,\eta)/\eta$, for
 $\sqrt{s}$ = 5500 GeV. The curves at $\eta=0$ correspond from
below to values of $\tau$ as in fig.~\ref{eps}.}}
\la{theta}
\end{figure}
The flow $\Theta(\tau,\eta)$, scaled by the Bjorken flow $\eta$, is
shown in fig.~\ref{theta}. The initial flow $\Theta=\eta$ is rapidly
accelerated at large $|\eta|$, due to increasing $\partial_\eta p/p$,
(=$\eta\slash\sigma^2$ for the gaussian parametrization) but this is
already in
the domain where the details of the model need not be correct. In the
relevant region near $\eta=0$ the effects are seen to be small, a few \%.

\begin{figure}[tb]
\vspace*{2cm}
\centerline{\hspace{-1.8cm}
\epsfxsize=10cm\epsfbox{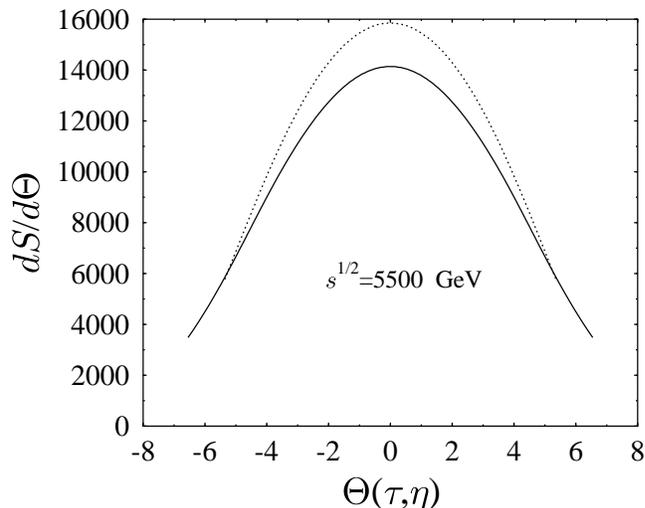}}
\vspace*{-6cm}
\caption[a]{{\small The total entropy per unit $\Theta$ between the
two thick flow lines in fig.~\ref{epsf} for $\sqrt{s}$ = 5500 GeV.
The dotted curve corresponds to $\tau=\tau_i$
and the solid one to $\tau=100\tau_i$.}}
\la{s}
\end{figure}

As discussed earlier, the total entropy between two flow lines,
eq.(\ref{stot}), is constant. However, the flow rapidity 
$\Theta(\tau,\eta)$ changes along
the flow lines changing also the rapidity interval between the flow lines.
As a result the entropy of given fluid element is shifted in rapidity and its
amount per unit flow rapidity is changed. With our initial conditions the
shift is always to larger rapidities --- higher pressure in the central region
accelerates the fluid towards the ends --- and the rapidity intervals between
the flow lines increase. The net result is that the entropy per unit
flow rapidity, which can be expressed as
\be
{dS(\tau,\eta)\over d\Theta}=\pi R_A^2 \tau s(\tau,\eta)
\cosh[\Theta(\tau,\eta)-\eta]{d\eta\over d\Theta}\Bigg|_{\rmi{fixed}\,\,\tau},
\ee
will decrease in the central region and the overall distribution will get
wider.

The result of our computation for $dS\slash d\Theta$ is shown in 
Fig.~\ref{s}. 
In $\eta$ the computation is extended between the two 
(arbitrarily chosen) thick flow lines in
Fig.~\ref{epsf}. The dotted curve shows the initial distribution
at $\tau_i=0.1$ fm and the solid curve the final distribution at
$\tau=100\tau_i$. Note that the end points move outwards due to
a combination of two effects: the flow lines bend outwards
(Fig.~\ref{epsf}) and the flow is accelerated relative to $\eta$
(Fig.~\ref{theta}).

To obtain a measurable distribution we should be able to treat
the hadronization and to fold the thermal motion of final particles with the
collective motion of the flow (after expressing the entropy in terms of
particle number densities). At present, we do not have a reliable
way to estimate the effects of hadronization but if it has any effects on flow
we would expect them to further widen the rapidity distribution as the
hadronization proceeds from the lower density fragmentation regions to the
central region.
Thermal folding involves flow
velocities effectively over two-to-three rapidity units.  It will lead to a
somewhat wider overall rapidity distribution but in the smooth central region
the effect of folding is small. We conclude that our result of Fig.~\ref{s}
gives the minimum change from the initial state
at $\tau_i$ to the rapidity distribution of final particles.

\section{Conclusions}

In the case of a fully developed rapidity plateau, there is a very simple
hydrodynamical scaling solution for the longitudinal expansion of QCD
matter produced in ultrarelativistic heavy ion collisions \cite{bj}.
In reality, there need not be a rapidity plateau,
and we have studied the longitudinal
flow using a  rapidity distribution of the initial energy density
obtained from computations in perturbative QCD. Then the rapidity
distribution is approximately gaussian but very broad, and thus
the deviations from the Bjorken flow are not very large. To analyze
the general features of the flow we also derived equations for the total
entropy and energy between two flow lines and gave approximate solutions
in the limit of broad gaussians for the central region.

The finite width of the rapidity region leads to a transfer of energy 
from the central region to larger rapidities. 
As a consequence the rapidity distribution
gets wider and the energy density 
in the central region decays faster than for a boost invariant 
flow. Correspondingly, the lifetime of the system (the time it takes to 
decrease from $\epsilon_i$ to $\epsilon_c$) 
decreases by about 10\%. A similar decrease would be caused by transverse
density gradients. On the other hand, dissipative effects would increase
the lifetime.

The longitudinal density gradient also leads to some but very small
acceleration of the longitudinal flow: the effect in the central region
is rather on the 1\% than 10\% level.  In the sense our results confirm the
assumptions in \cite{vjmr}, called the frozen motion model in
\cite{sx}, that in
estimating thermal effects in the central
region it is reasonable to assume that at LHC energies the longitudinal
flow velocity scales even when the density distributions are not boost
invariant.

\end{document}